# Exploring the Interplay Between Formation Mechanisms and Luminescence of Lignin Carbon Quantum Dots from Spruce Biomass

*Jelena Papan Djaniš[†,‡,*], Maja Szymczak[†], Jan Hočevar[†], Jernej Iskra[†], Boštjan Genorio[†], Darja Lisjak[⊥], Lukasz Marciniak[†], Karolina Elzbieciak-Piecka[ℓ,*]*

[†] Faculty of Chemistry and Chemical Technology, University of Ljubljana, 1000 Ljubljana, Slovenia

[‡] Centre of Excellence for Photoconversion, Vinča Institute of Nuclear Sciences, National Institute of the Republic of Serbia, University of Belgrade, 11351 Belgrade, Serbia

[⊥] Department for the Synthesis of Materials, Jožef Stefan Institute, 1000 Ljubljana, Slovenia

[ℓ] Institute of Low Temperature and Structure Research Polish Academy of Sciences, Wroclaw

*Corresponding authors: jelena.papandjanis@fkkt.uni-lj.si and k.elzbieciak@intibs.pl

ABSTRACT: This study investigates the intricate relationship between the formation mechanisms and luminescent properties of lignin-derived carbon quantum dots (LG-CQDs) synthesized from spruce biomass by hydrothermal treatment. A comprehensive understanding of LG-CQD structure and its photoluminescence requires insights into the native architecture of lignin and the distribution of its acidolysis-derived fragments. Research showed how these lignin-derived units interact with dopant molecules in three different approaches during synthesis, contributing to core and surface structures that govern the optical behavior. Our findings reveal a clear correlation between structural features and luminescent properties, emphasizing the role of surface chemistry in tuning emission characteristics. These insights provide a foundation for the rational design of LG-CQDs with tailored luminescent properties, advancing their potential applications in sustainable optoelectronics, sensing, and bioimaging.

**Introduction**

In an era when fossil fuels enormously influence the pollution of the global environment, it is more than desirable to use abundant materials and cost-effective methods to produce new materials. One of the most abundant materials on Earth is lignocellulosic biomass, which consists of three main components: cellulose, hemicellulose, and lignin [1]. While the first two components are widely used in many areas, only a small percentage of lignin, the waste material of the paper industry, is used commercially. The rest is mainly used as low-grade fuel [2]. The limited utilization of lignin is primarily due to its complex and heterogeneous structure, which varies significantly between plant species. Nevertheless, lignin offers a promising and versatile platform for the development of multifunctional materials, providing both scientific challenges and opportunities [3]. Lignin is the only natural polymer with a rich aromatic structure consisting of three aromatic alcohols: *p*-hydroxyphenyl (H), guaiacyl (G), and syringyl (S) [4]. As recently demonstrated, it possesses luminescent properties that can be enhanced by transforming lignin into lignin carbon quantum dots (LG-CQDs), spherical or quasi-spherical carbon nanoparticles with sizes below 10 nm [5], [6]. Since 2016, when the first study on LG-CQDs was published [6] numerous notable publications have emerged exploring the use of lignin as a precursor for LG-CQDs. These studies have focused primarily on tuning luminescence properties, such as emission color and quantum yield, as well as investigating the application of LG-CQDs in sensing technologies [7], [8], [9], [10]. For instance, lignin has been doped with mild organic acids such as 4-aminobenzoic acid, benzenesulfonic acid, 4-aminobenzenesulfonic acid, and 2,4-diaminobenzenesulfonic acid, enabling a detailed investigation of the fluorescence mechanisms responsible for multicolor-emitting CQDs [5]. In another approach, doping lignin with small nitrogen-containing molecules — such as ethylenediamine co-doped with magnesium — resulted in CQDs exhibiting high sensitivity to pH variations [6]. Additionally, LG-CQDs doped with *m*-phenylenediamine have been evaluated as formaldehyde sensors, while doping with *o*-aminobenzenesulfonic acid produced CQDs with high sensitivity toward $Fe^{3+}$ ions. Doping with aminophenylboronic acid enabled the development of a selective sensor for $Cr^{6+}$ ions. [8], [11], [12]. Yhu et al. demonstrated that by varying the ratio of nitrogen and sulfur atoms introduced as substituents on aromatic dopants, the emission color of carbon quantum dots can be precisely tuned across the visible spectrum – from blue and green to yellow and red [13]. In the broader context of luminescence tuning, it is important to emphasize that lignin is typically doped with electron-donating substituents such as −$NH_2$, −NHR, −$NR_2$, −OH, −OR, or −CN. These electron-donating groups, in conjunction with aromatic rings, contribute to an extended π-conjugated system through delocalization of electrons. This enhanced π-electron conjugation effectively increases the luminescence efficiency of the resulting CQDs [14].

It is important to recognize that the intrinsic structure of lignin plays a critical role in the formation of LG-CQDs [15]. Typically, LG-CQDs are synthesized via a two-step process: an initial acidolysis step followed by a hydrothermal reaction. The primary objective of acidolysis is the depolymerization of the amorphous, three-dimensional polymer network of lignin.

However, this depolymerization leads to a heterogeneous mixture of fragmented polymeric units of varying sizes. These size distributions can be partially narrowed through post-treatment techniques such as centrifugation, filtration, and dialysis [15]. During the subsequent hydrothermal step, these lignin-derived fragments, in the presence of selected dopants, undergo further transformation to form complex nanocarbon structures. Despite employing identical synthesis parameters, including dopant type, solvents, reaction conditions, and purification protocols, the resulting LG-CQDs often exhibit variations in their degree of carbonization and surface functionalization. These structural differences significantly influence their optical behavior, particularly their luminescence properties [16].

A comprehensive understanding of the structure and luminescent properties of LG-CQDs requires in-depth knowledge of both the native lignin architecture and the distribution of lignin-derived fragments produced during acidolysis. Equally important is elucidating the mechanism of LG-CQD formation, wherein smaller lignin-derived units interact with dopant molecules. Additionally, gaining insight into how surface chemistry influences the optical behavior of these nanostructures is a critical step toward expanding their potential applications.

Our research aims to address these fundamental questions. In this study, we utilized lignin isolated from spruce biomass and doped it with *m*-aminophenylboronic acid. The systematic investigation of the effect of varying hydrochloric acid (HCl) concentrations, introduced alongside the organic dopant, on the synthesis process, structural characteristics, and luminescent properties of the resulting LG-CQDs has been performed.

**Materials and methods**

Kraft lignin was provided by Lignocity from Sweden, *m*-aminophenylboronic acid was purchased from Fluorochem, while hydrochloric acid (37% aq. sol.) was purchased from Pregl chemicals, and NaOH (p.a.) was purchased from SigmaAldrich.

*Synthesis of the LG-CQDs*

LG-CQDs were produced using a two-step, eco-friendly hydrothermal synthesis [11], [17]. In the first approach, 0.3g of lignin and 30 ml of DI water were mixed and sonicated in the water bath for 10 minutes. After sonication, 0.3 g of *m*-aminophenylboronic acid was added under constant stirring in the previously mentioned mixture. The mixture was then heated at 90 °C for 1 h with constant stirring. After 1h, the reaction was stopped and cooled to room temperature and filtered through a nylon filter (0.45 µm), while the filtrate was transferred into a Teflon-lined autoclave (50 ml) and heated at 200 °C for 12 h. The obtained solution was cooled to room temperature, sonicated for 10 minutes, and further filtered through a microporous filter (0.22 µm). The remaining filtrate was dialyzed in a dialysis bag (500-1000 Da) against DI water for 2 days. The as-dialyzed CQDs aqueous solution was stored as a liquid at 4 °C or freeze-dried under −60 °C for 2 days and characterized as a powder. The sample was referred to as *N, B CQDs*. The initial concentration of *N, B CQD* after purification was 0.03 mg ml$^{-1}$.

In the second approach, 0.3g of lignin and 30 ml of DI water were mixed and sonicated in the water bath for 10 minutes. After sonication, 20 μl HCl was added and then stirred and heated at 60 °C for 30 minutes. The addition of 0.3 g of *m*-aminophenyl boronic acid and all the next steps of reaction and purification were the same as in the previous procedure. The sample was referred to as *N, B CQD 20 μl HCl*. The initial concentration of *N, B CQD 20 μl HCl* after purification was 0.19 mg ml$^{-1}$.

In the third approach, the same protocol described before was used with 1000 μl HCl. The sample was referred to as *N, B CQD 1000 μl HCl*. The initial concentration of *N, B CQD 1000 μl HCl* after purification was 0.88 mg ml$^{-1}$.

*Characterization*

Infrared spectra of the synthesized carbon quantum dots were recorded between 4000 cm$^{-1}$ and 400 cm$^{-1}$ using a Bruker Alpha-II (FT-IR spectrophotometer with ATR module). Raman spectra were recorded with Bravo Handheld Raman Spectrometer, Bruker Optic.

Electro-kinetic measurements (zeta potential) and dynamic light scattering (DLS) analysis of CQDs dispersed in double deionized water with a concentration of 1 mg ml$^{-1}$ were performed using a Litesizer 500 (Anton Paar). The pH was adjusted with HCl and NaOH solutions (0.1 or 1 M).

A total of 40 mg of dry, pre-isolated lignin was dissolved in 0.5 mL of deuterated DMSO (d-DMSO). The solution was then filtered and transferred to an NMR tube for analysis. 2D-HSQC NMR spectra were recorded at 25 °C using a Bruker Avance III 500 spectrometer. The following acquisition parameters were used: F2 range = 10 to 0 ppm, F1 range = 158 to –8 ppm, number of scans (ns) = 24, dummy scans (ds) = 16, number of increments (ni) = 256, relaxation delay (d1) = 1.47 s, and pulse program = hsqcetgpsi2. The spectra were processed with Bruker TopSpin 4.1.1 and MestReNova software.

X-ray photoelectron spectroscopy (XPS) measurements were conducted using a Versa Probe 3 AD (PHI, Chanhassen, USA) equipped with a monochromatic Al *Kα* X-ray source. The source operated at an accelerating voltage of 15 kV and an emission current of 13.3 mA. Powder samples were mounted on double-sided Scotch tape and positioned at the center of the XPS holder. Spectra were acquired for each sample over a 200 μm × 200 μm analysis area, with the charge neutralizer activated during data collection. Survey spectra were measured using a pass energy of 224 eV and a step size of 0.8 eV. High-resolution (HR) spectra were recorded with a pass energy of 27 eV and a step size of 0.05 eV. To ensure high-quality spectral data with a good signal-to-noise ratio, at least 10 sweeps were performed for each measurement. The energy scale of the XPS spectra and any possible charging effects were corrected by referencing the C=C peak in the C 1s spectrum of the carbon support, with a binding energy (BE) of 284.5 eV. Spectral deconvolution was carried out using MultiPak 9.9.1 & Casa XPS software, and Shirley background correction was applied to all spectra.

LG-CQDs were analyzed with a transmission electron microscope (TEM, Jeol 2100, Tokyo, Japan). Colloids were drop-deposited on the Cu-supported TEM grid and left to dry.

Excitation and emission spectra and luminescence decay profiles of liquid CQDs were recorded using FLS 1000 Fluorescence Spectrometer from Edinburgh instruments equipped with a R928P side window multiplier tube from Hamamatsu as a detector as well as a 450 W halogen lamp and picosecond pulsed light source AGILE from Edinburgh instruments as the excitation sources. The temperature during the measurements was externally controlled using a Quantum Northwest TC1 Temperature Controller with associated Peltier-controlled cuvette holder. The pH value of liquid CQDs was controlled and stabilized to pH=7 before measurements using a Seven Compact S210 pH meter from Mettler Toledo.

**Results**

**Structure and the surface chemistry of N, B CQDs**

Nanoparticles with a core diameter of approximately 10 nm were synthesized by using a classical two-step reaction method, which is well established in the literature and involves mild acidolysis with *m*-aminophenylboronic acid [12]. These nanoparticles formed predominantly agglomerates in aqueous solution with a size from 180 to 1000 nm and an average hydrodynamic diameter of 290 ± 54 nm (Figure 1a), d)). Pretreatment of the lignin with 20 μl HCl led to the formation of smaller carbon quantum dots decorated with an amorphous lignin network packed solely in water, with average diameter of 1.7 ± 0.3 nm or in agglomerates ranging from 50 to 290 nm, with an average diameter of 118.6 ± 39.5 nm (Figure 1b), e)). The addition of 1000 μl HCl in a pretreatment process led to the formation of a higher concentration of LG-CQD, with a homogeneous size of a few nanometers (Figure 1 c). Two distinct populations of nanoparticles were observed in water: individual LG-CQD with an average size of 1.32 ± 0.80 nm and agglomerates of LG-CQD with an average diameter of 127 ± 54 nm, with agglomerates ranging in size from 18–370 nm (Figure 1f). Contrary to previous reports [8], [12], [18], which describe the synthesis of CQDs from lignin, our results on kraft spruce lignin indicate that a pretreatment with a HCl is essential. This acid pretreatment initiates partial depolymerization of the lignin structure, rendering it more reactive and suitable for subsequent functionalization with *m*-aminophenylboronic acid.

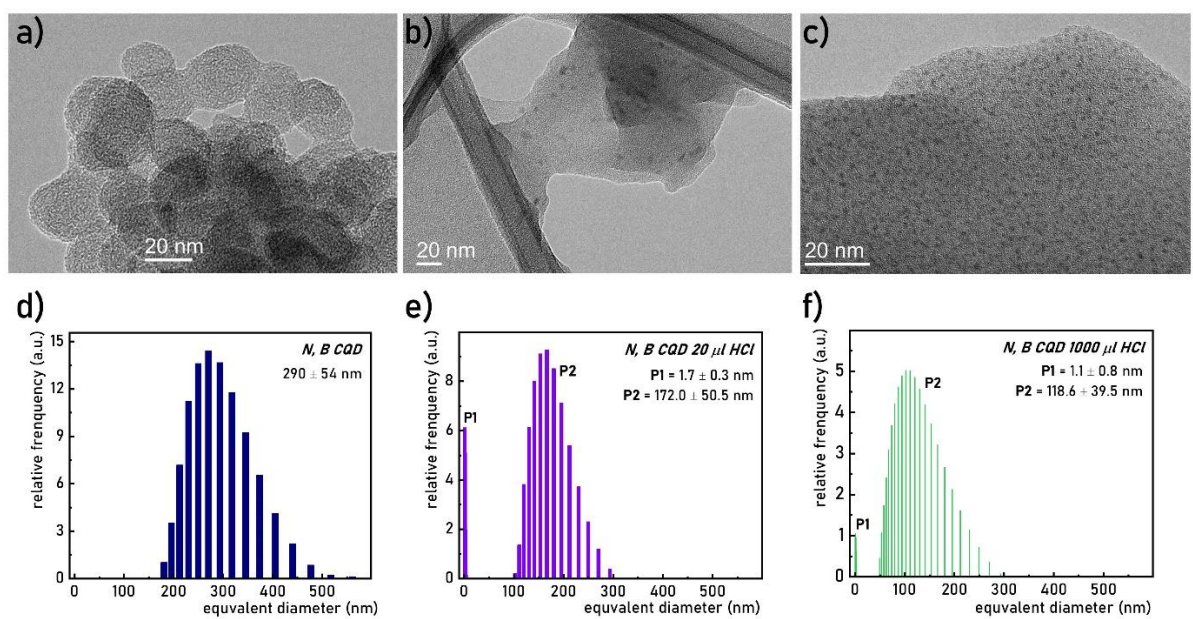

**Figure 1**: Representative TEM images of the a) N, B CQD b) N, B CQD 20 μl HCl, c) N, B CQD 1000 μl HCl, and corresponding number-weighted distributions of the hydrodynamic diameters of d) N, B CQD e) N, B CQD 20 μl HCl, f) N, B CQD 1000 μl HCl.

The successful reaction between lignin units and m-aminophenylboronic acid can be confirmed using FTIR and zeta potential measurements, both of which allow monitoring of changes in the surface chemistry of the resulting LG-CQDs. FTIR analysis of spruce-derived lignin revealed characteristic bands associated with aromatic polyphenolic structures (Figure 2a). A band at 1034 cm$^{-1}$ can be assigned to C-H aromatic vibrations, 1085 and 1269 cm$^{-1}$ belong to C-O vibrations, 1217 cm$^{-1}$ to G unit, 1511 cm$^{-1}$ to aromatic C=C, 1601 cm$^{-1}$ to aromatic C-O, while 1714 cm$^{-1}$ can be assigned to C=O bands. The band at 2942 cm$^{-1}$ is associated with C–H stretching, and a broad absorption around 3400 cm$^{-1}$ is characteristic of O–H groups [19]. After reaction with *m*-aminophenylboronic acid, new absorption bands appear at 1039 and 1087 cm$^{-1}$, which could be associated with B-O-H and B-O vibrations, while an additional band at 3217 cm$^{-1}$ belongs to the amino group (Figure 2a).

A closer look at zeta potential reveals that the use of HCl in a pretreatment affects the surface chemistry of the synthesized *N, B CQD*s (Figure 2b) [20]. The surface of *N, B CQD*s has a negative charge in the pH range 2.5-10, which means that the dominant functional groups are –OH. In samples pretreated with 20 μl of HCl, the isoelectric point is close to pH 2.5, indicating an increased availability of surface amine groups. Partial protonation of amino groups in this pH range reduces the absolute value of the zeta potential [21]. The addition of 1000 μl HCl in the third approach resulted in a shift of the isoelectric point to pH=6.5. Such a shift is typically associated with amino-containing surface functionalities, as observed in the case of *p*-phenylenediamine groups [22]. Thus, the addition of 1000 μl HCl resulted in an increased concentration of amino groups on the surface of *N, B CQD 1000 μl HCl* compared to *N, B CQD* and *N, B CQD 20 μl HCl*.

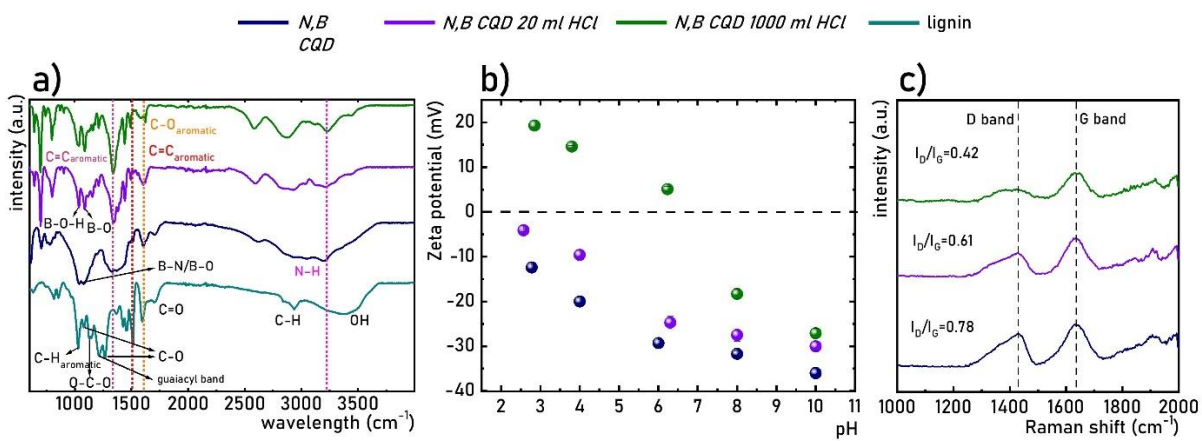

**Figure 2**: a) FTIR spectra of the lignin from spruce and the powders of *N, B CQD*s, *N, B CQD 20 μl HCl* and *N, B CQD 1000 μl HCl*, b) Zeta potential of colloids of LG-CQD, c) Raman spectra of colloids of LG-CQD.

Raman spectra of CQD exhibit two characteristic bands, where the D band is mainly present at ~1350 cm$^{-1}$, associated with the defects in the graphite lattice caused by sp$^3$ carbon structure, and the G band is present at ~1580 cm$^{-1}$, corresponding to the graphitized sp$^2$ carbon structure [18]. Figure 2c presents the Raman spectra of the synthesized LG-CQDs, where the D band is near 1400

cm$^{-1}$, and the G band around 1630 cm$^{-1}$. Both values of D and G bands are in accordance the literature [15]. The intensity ratio of the D to G bands ($I_D/I_G$) serves as an indicator of the degree of structural order within the carbon framework. For the as-prepared *N, B CQD*s (without pretreatment), the $I_D/I_G$ ratio was 0.78. This ratio decreased to 0.61 for the sample pretreated with 20 μl of HCl, and further to 0.42 with 1000 μl of HCl in the pretreatment phase. These results indicate that acid pretreatment enhances the graphitization of the CQDs, promoting the formation of more ordered sp² carbon domains [23].

XPS analysis was performed to gain a more detailed understanding of the surface chemistry of the synthesized CQDs. The survey spectra revealed five characteristic peaks: C 1s, O 1s, N 1s, B 1s, and Cl 2p at binding energies of 284.5 eV, 531.7 eV, 400.3 eV, 190.9 eV, and 197.5 eV, respectively (Figure S1). Quantitative XPS analysis (Table 1) indicates that the addition of HCl during synthesis increases Cl, B, and N surface concentrations. Furthermore, the addition of HCl alters the nitrogen composition, increasing the contribution of graphitic nitrogen and promoting a higher degree of graphitization. These findings align with our Raman spectra (Figure 2c) and previous studies [23].

**Table 1**: The quantitative XPS analysis of elements in CQDs

| Sample | C (at %) | O (at %) | N (at %) | B (at %) | Cl (at %) | N-amino | N-graphitic |
|---|---|---|---|---|---|---|---|
| *N, B CQD* | 69.88 | 15.75 | 5.27 | 2.78 | 0.34 | 0.6 | 0.4 |
| *N, B CQD 20 μl HCl* | 68.16 | 16.18 | 6.60 | 3.47 | 2.93 | 0.45 | 0.55 |
| *N, B CQD 1000 μl HCl* | 68.92 | 14.83 | 6.75 | 3.96 | 3.42 | 0.43 | 0.57 |

To analyze surface functionalities, a high-resolution C 1s spectrum was recorded and deconvoluted into distinct components. For an accurate fitting, C 1s core-level spectra of lignin and 3-aminophenylboronic acid, the starting materials, were also recorded (Figure 3a-b). The C 1s peak of LG-CQDs was fitted with seven components corresponding to C=C, C-C (C-H), C=O, C-O, C-N, and C-B bonds at 284.5, 284.9, 286.2, 288.9, 289.8, 285.7, and 284.0 eV, respectively (Figure 3c-e). Binding energies and full widths at half maximum (FWHM) values were determined based on reference materials. The proportion of *sp²*-hybridized carbon increased by 124% (*N, B CQD*), 101% (*N, B CQD 20 μl HCl*), and 121% (*N, B CQD 1000 μl HCl*) compared to lignin (Table S1-S4).

Additionally, Figure 3 shows a decreasing trend in C-O functional groups as follows: *N, B CQD > N, B CQD 20 μl HCl > N, B CQD 1000 μl HCl*. Interestingly, no COOH groups were detected on the surfaces of either lignin or CQDs.

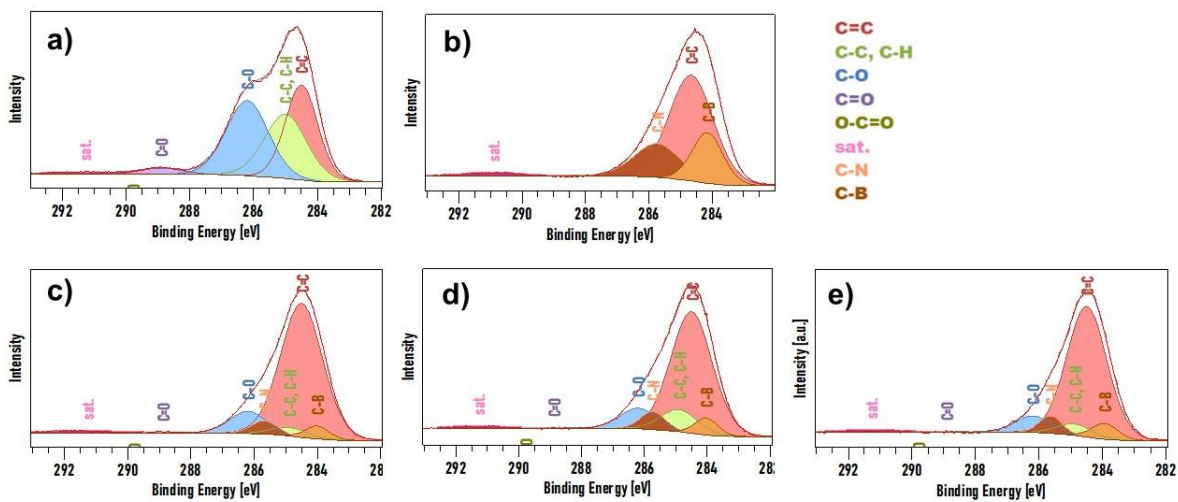

**Figure 3**: C1s Core level XPS spectra of starting compounds – a) Lignin and b) 3-aminophenylboronic acid and c) *N, B CQD* d) *N, B CQD 20 μl HCl*, e) *N, B CQD 1000 μl HCl*

For a deeper understanding of nitrogen functionalities, high-resolution N 1s spectra were recorded and fitted with two components at 399.8 eV and 400.8 eV, corresponding to amino nitrogen and graphitic nitrogen, respectively (Figure S2) [24], [25]. With a higher ratio of added

HCl, the structure of N, B CQDs exhibited increased incorporation of nitrogen within an aromatic core structure, labelled as graphitic nitrogen (Table 1). Based on these results, we proposed a surface model for our CQDs as in Figure 4.

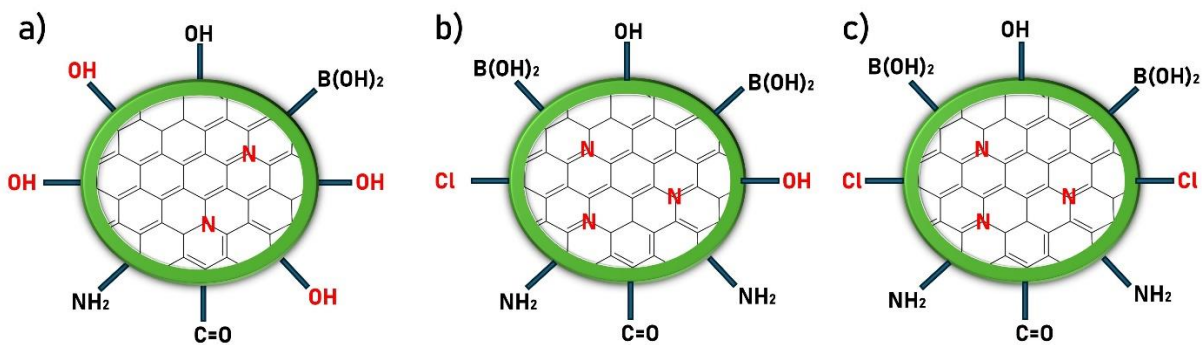

**Figure 4**: Schematic representation of the surface of a) *N, B CQDs*, b) *N, B CQD 20 μl HCl*, and c) *N, B CQD 1000 μl HCl*.

**NMR study and mechanism approach**

A comprehensive two-dimensional heteronuclear single quantum coherence (2D HSQC) NMR spectroscopy analysis was performed to investigate the structural transformations of Kraft spruce lignin, its acidolysis residues, and the LG-CQDs synthesized from them. The 2D HSQC spectra are shown in Figure S3. Detailed assignments of the $^1$H-$^{13}$C chemical shift can be found in the Supporting Information (Table S5).

Spruce lignin consists mainly of guaiacyl (G) and *p*-hydroxyphenyl (H) units, while syringyl (S) units are almost undetectable (< 1.5%). The HSQC spectrum of spruce kraft lignin shows a large number of signals in the side chain region (50–90 ppm/3.2–5.0 ppm) corresponding to important structural features such as methoxyl (-OCH$_3$) groups, β-aryl ether linkages (A), phenylcoumaran (B) structures and resinol (C) linkages (Figure S3) [26]. In addition, the 2D-HSQC NMR spectrum of spruce kraft lignin shows the presence of a signal associated with the aromatic moiety of ferulic acid.

As can be seen in Figure 5a, acidolysis leads to a significant reduction in the intensity of the side chain signals corresponding to structural units A, B and C, indicating extensive de-etherization of kraft lignin [27]. The number of propanoid bonds between the aromatic units is reduced from the original 188 to only 0–8 per 1000 aromatic units (Figure 5a). Given the very low number of propanoid bonds per 1000 aromatic units after hydrolysis or hydrothermal synthesis of CQDs, the relative proportions of each bond type are less reliable. Interestingly, when a small amount of acid was used (20 µl HCl), a slight increase in the number of bonds was observed, although the difference is small in comparison to the parent lignin. The effect of a larger amount of HCl is very pronounced, as the A, B, and C bonds disappear. Similarly, the signals corresponding to the oxidized form of ferulic acid also disappear [28].

These results indicate that the sequential two-step synthesis effectively cleaves the ether bonds, leading to significant structural changes within the lignin scaffold. In particular, a significant shift in the ratio of aromatic units G/H is observed during acidolysis and hydrothermal synthesis of CQDs, which is indicative of a demethylation process, as shown in Figure 5b. The ratio of the integral in the HSQC NMR spectrum between the methoxy region and the total aromatic signal decreases from 1.6 in the original spruce kraft lignin to 0.2 in the resulting N,B-CQDs. This reduction is even more pronounced when HCl is used in the synthesis.

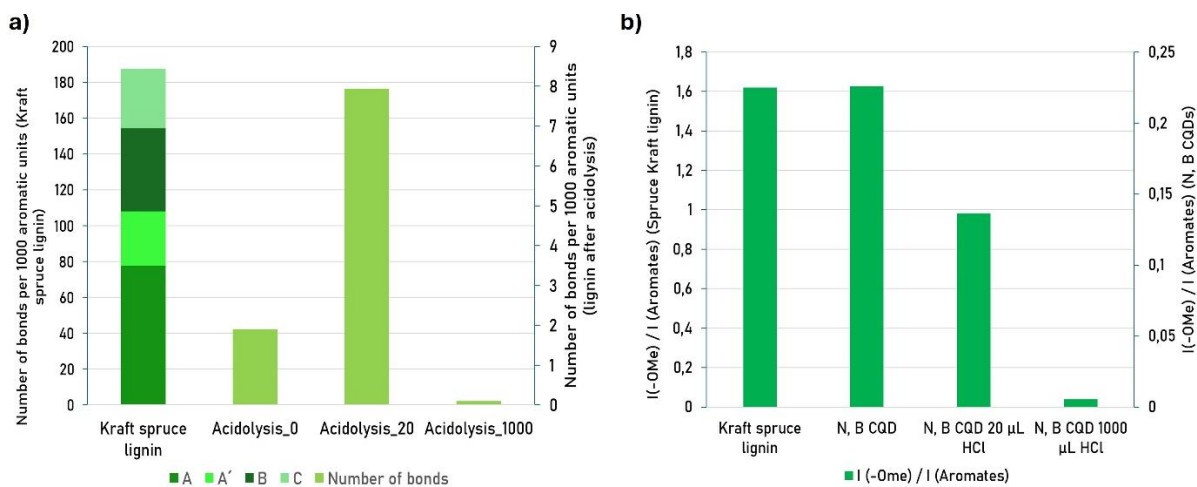

**Figure 5**: a) Ratio of bond types between units in kraft spruce lignin and the corresponding number of bonds per 1000 aromatic units after acidolysis, and b) ratio of the HSQC integral for the methoxy group domain to the total aromatic domain in pine kraft lignin or N,B-CQDs.

In addition, the HSQC spectra of samples after acidolysis with HCl show a shift of the aromatic signals to higher chemical shift values (Figure 6). This is consistent with the XPS analysis, which confirms the Cl atom's binding to the aromatic lignin rings. Moreover, a decrease in the ferulic acid signal is observed, accompanied by the appearance of a signal corresponding to the oxidized form (pCA$_{2,6}$). These shifts become more evident when a larger amount of HCl (1000 µl) is used. In this case, the aromatic units exhibit partial shift (indicating partial functionalization), such as the units labelled G$^0_2$, G$^0_5$, G$^0_6$ and H$^0_{2,6}$, while a larger proportion of the units exhibit stronger shift to higher ppm values, indicating a higher degree of functionalization. No significant shifts in the aromatic region of the NMR spectra are observed during the hydrothermal synthesis process, indicating that the primary functionalization of the aromatic units of lignin occurs during the acidolysis (Figure 6). The shifts observed in the aromatic N,B region of the CQDs are consistent with those reported for simpler aromatic compounds functionalized with boron and chlorine.

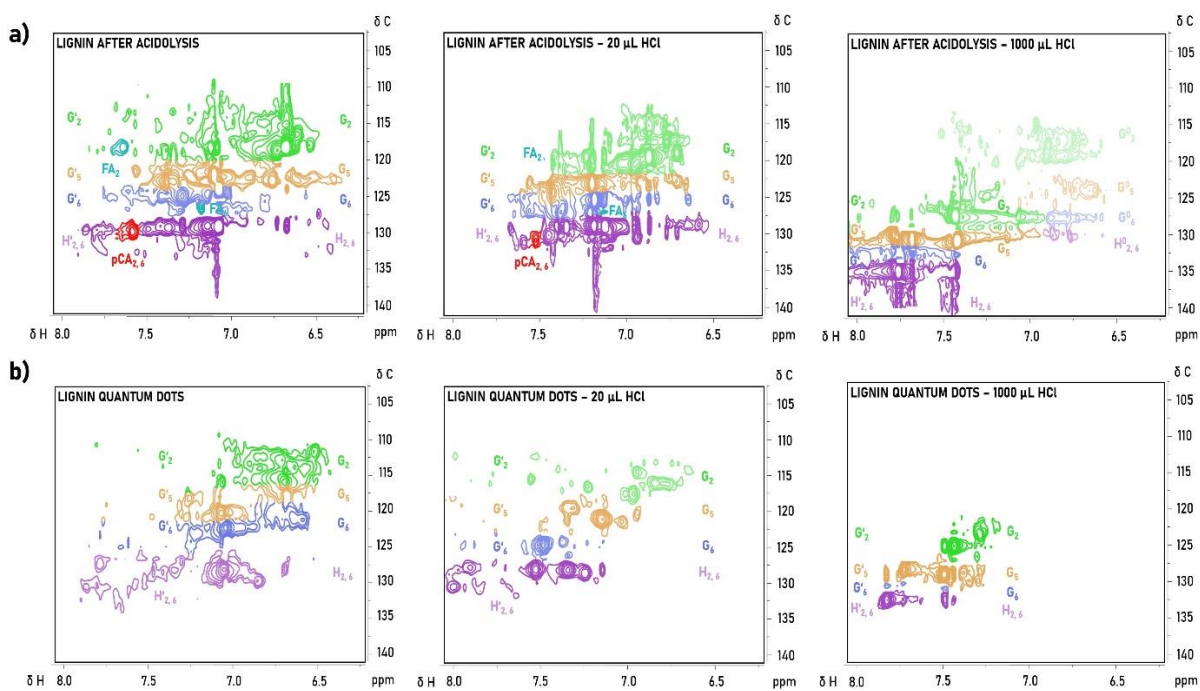

**Figure 6**: 2D-HSQC NMR spectra of aromatic area of a) acidolysis components, and b) CQDs.

Based on the results presented above, we propose the following mechanism for the formation of N, B-co-doped CQDs from spruce kraft lignin via a two-step approach. In the first step, the kraft lignin is de-etherified by acidolysis, resulting in monomeric lignin units. At the same time, N- and B-doping is achieved by incorporating 3-aminophenylboronic acid. This reagent has a dual function: it promotes the cleavage of ether bonds and simultaneously introduces N and B atoms into the monomeric lignin units [29]. The concentration of HCl used in the pretreatment influences the efficiency of both processes—ether bond cleavage and doping.

In the subsequent hydrothermal synthesis step, the doped monomeric lignin units serve as precursors for CQD formation. This process involves simultaneous dehydration and condensation reactions that promote the formation of π-π interactions between the modified aromatic units, ultimately leading to CQD synthesis [12].

Slightly distinct structural features of the N, B-CQDs obtained with 20 µl HCl are reflected in the increased number of bonds between the units, which indicates the change in their bonding patterns. In this case, the lignin structure is less disrupted compared to higher HCl concentrations. One of the possible explanations of this effect could be that in the complex mixture during acidolysis, there is a competition between lignin and *m*-aminophenylboronic acid, where both react with HCl. The usage of a smaller amount of HCl might lead to a partial cleavage of lignin bonds and direct functionalization of *m*-aminophenylboronic acid, while a higher amount of acid was enough for total cleavage as well as functionalization of dopant acid. These findings are considered together with the results of the XPS analysis, where *N, B CQD 20µl HCl* had a stronger oxidation of the lignin units and a slightly lower functionalization than with a higher amount of HCl, which ultimately leads to different structural properties of the resulting CQDs.

**Luminescence of CQDs**

Due to their unique optical properties, CQD exhibit considerable potential for a wide range of applications. To evaluate this potential and support further development, the obtained CQDs materials were subjected to spectroscopic analysis. As is generally known, the mechanism of luminescence of CQD is influenced by many factors. The main mechanisms considered are based on band gap transitions of conjugated π-domains, surface defects and molecular fluorescence[30], [31], [32], [33]. However, the exact origin of luminescence must be individually assessed for each CQD system, considering variables such as the nature of the precursor, the synthesis route, the resulting carbon structure, the surface composition, the incorporation of dopants, etc. [34], [35], [36].

In the case of the analyzed materials, the most plausible explanation for the occurring luminescence is the presence of surface states (Figure 7a). As illustrated in the energy-level diagram, the main emission band originates from the containing conjugated pair of π-electrons carbon core, what could be explained as an electron transfer from the lowest unoccupied molecular orbital (LUMO) to the highest occupied molecular orbital (HOMO) [37], [38], [39]. However, acidolysis-induced modifications in surface chemistry, particularly in N,B-CQDs pretreated with 20 µl of HCl — are associated with a reduction in the energy gap, which enhances emission from surface states, particularly in the green spectral region. Further evidence for the role of surface defects in this process is provided by the excitation-dependent emission behaviour of the CQDs [40], [41] (Figure 7b-d). As can be seen in Figure 7b, optical excitation in the spectral range between 260–300 nm produces an emission band centred at 375 nm, with increasing intensity and spectral broadening. However, upon 320 nm excitation line an additional emission band appears with a maximum at 412 nm. The 340 nm excitation induces both a shift in the emission maximum to 421 nm and a corresponding change in spectral shape. This band increases in intensity for excitation at 360 nm and 380 nm. However, in the case of excitation at 400 nm, a second band with lower intensity was observed in the emission spectra, together with the band centered at 425nm. With a change in the excitation wavelength, i.e. from 420 nm to 480 nm, only one band with a maximum at 540 nm can be recognized, corresponding to the green spectral range. A similar excitation-dependent pattern is observed for N,B CQD 1000 µl HCl (Figure 7d), with the notable difference having a stronger 418 nm emission alongside the 375 nm band in the 260–300 nm excitation range. On the other hand, with 320 nm excitation both of these bands remain with lower intensity. In contrast, the N,B CQD sample pretreated with 20 µl HCl exhibits a distinct emission profile. In the excitation range of 260 nm-320 nm, only one band with low-intensity is observed at 418 nm, while the 375 nm band is absent. At 400 nm excitation, in addition to the emission band with the maximum at 425 nm, a clearly pronounced second emission band with the maximum at around 505 nm was observed. With excitation in the 420 -480 nm range, the emission spectra show variations in the band shape, with the maxima shifting around 515 nm depending on the excitation wavelength. The above observations are supported by the excitation spectra of the analysed materials measured with the emission monitored at 425 nm (Fig. 7e). A dominant, broad band centred at 370 nm was observed in all spectra, together with bands of lower-intensity at approximately 270 nm and 320 nm for *N,B CQD* and at around 260 nm and 300 nm for *N,B CQD 20 μl HCl* and *N,B CQD 1000 μl HCl*. These excitation features explain the presence of emission bands in the UV-range. Based on the performed excitation-emission mapping, the optimal excitation wavelength of 400 nm was selected for further investigation as it provides strong emission and is suitable for luminescence decay measurements. Comparative analysis of the normalized emission spectra shows that the *N,B CQD* and *N,B CQD 1000 μl HCl* samples exhibit very similar emission characteristics. However, the *N,B CQD 20 μl HCl* sample shows a significant increase in the relative intensity of the second emission band, with a maximum at 503 nm. The analysis of the luminescence decay profiles of the investigated materials revealed their singleexponential character (Figure 7h). The determined luminescence lifetime values for *N,B CQD* and *N,B CQD 1000 μl HCl* are very similar and equal to 2.88 ns and 2.71 ns, respectively. However, for *N,B CQD 20 μl HCl* a slight elongation of the luminescence decay profile to the lifetime value to 3.18 ns was observed.

   Based on the spectroscopic analysis performed, it is clear that the luminescence properties of the *N,B CQD 20 μl HCl* differ from those of the other materials, which in turn show a great similarity in the luminescence response, although the *N,B CQD 1000 μl HCl* was subjected to an acidolysis process with a large amount of HCl, which altered the surface composition. To understand these differences, one must consider the structural and compositional data discussed previously.

By examining the XPS results, a significant change in surface composition can be observed between *N, B CQD*, *N, B CQD 20 μl HCl,* and *N, B CQD 1000 μl HCl*. Firstly, there was an increase in the number of nitrogen atoms originating from both amine groups and the graphitic nitrogen structure, as well as oxygen and boron. Chlorine groups are detected on the surface of *N, B CQD 20 μl HCl* and *N, B CQD 1000 μl HCl*. It is also worth noting the 2D-HSQC NMR spectra for the investigated materials (Figure S3 and Figure 6), where the clear differences can be seen both in the presence of specific subunits or functional groups on the spectrum, as well as in the shifts between them. This is particularly true for *N, B CQD 20μl HCl*, where there is an increased number of bonds after acidolysis compared to *N, B CQD* and *N, B CQD 1000μl HCl*. This significant change in the aromatic composition probably affects the structure of the carbon core and could be responsible for the observed differences in luminescence behaviour.

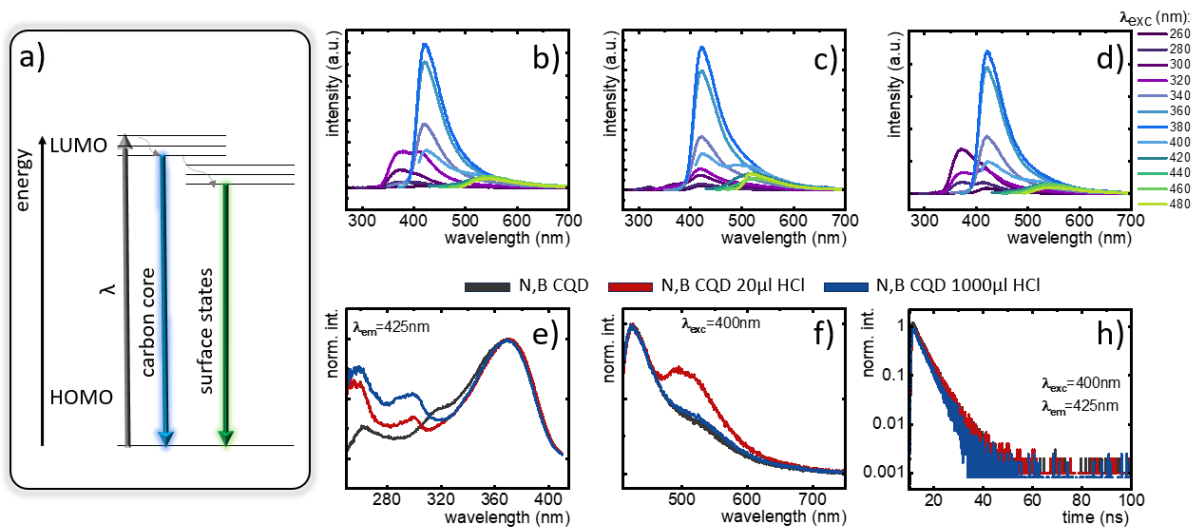

**Figure 7**: a) Simplified diagram of the luminescence mechanism in *N,B CQDs*; emission spectra of: b) *N,B CQD,* c) *N,B CQD 20μl HCl,* and d) *N,B CQD 1000μl HCl*; e) comparison of the room temperature excitation spectra ($\lambda_{em}$=425nm) and e) emission spectra upon $\lambda_{exc}$=400nm; f) luminescence decay curves ($\lambda_{em}$=425nm) -h) of the synthesized CQDs.

**Conclusions**

In this study, spruce lignin was used for the synthesis of lignin carbon quantum dots (LG-CQDs) through a stepwise process involving acidolysis and hydrothermal reactions. Three different syntheses were performed using *m*-aminophenylboronic acid as a dopant and different concentrations of HCl to promote acidolysis. The resulting LG-CQDs exhibited distinct surface chemistries, with the addition of HCl facilitating the incorporation of N, B, and Cl atoms on the surface and influencing the nitrogen content in the aromatic core. Our findings highlight the crucial role of the acidolysis phase in tuning the luminescent properties of the LG-CQDs, as well as the concentration of HCl. During acidolysis smaller amount of (20 µl) had a stronger oxidation of the lignin units and a slightly lower functionalization than with a higher amount of HCl (1000 µl). The usage of a smaller amount of acid led to the formation of new reaction paths, which resulted in an increase in the bond number between aromatic units after acidolysis, in comparison with approaches where no HCl is used and with the usage of a higher amount of HCl. We assume that a smaller amount of HCl had a competitive reaction between lignin and *m*-aminophenylboronic acid, where it partially reacted with both and caused partial cleavage of lignin units and partial functionalization of *m*-aminophenylboronic acid. Surface functional groups (such as -OH, $NH_2$, -Cl, and $B(OH)_2$) were found to directly influence the luminescence properties of the LG-CQDs. Excitation-emission mapping revealed that the optimal excitation wavelength of 400 nm produced strong emission, which was suitable for further investigation and measurements of luminescence decay. Comparative analysis of the emission spectra showed that the *N, B-CQD* and *N, B-CQD 1000 µl HCl* samples exhibited similar emission characteristics, while the *N, B-CQD 20 µl HCl* sample displayed a significant increase in the intensity of the second emission band, with a peak at 503 nm. This increased emission is attributed to the changes in surface chemistry caused by acidolysis, which led to a reduction in the energy gap between the HOMO-LUMO orbitals and increased emission from surface states in the green spectral region.

**Supporting Information**:

**Figure S1**: Survey spectra with quantification of a) Lignin, b) 3-aminophenyl boronic acid, c) N, B CQD, d) N, B CQD 20 μl HCl, e) N, B CQD 1000 μl

**Figure S2**: N1s core level spectra with deconvolution of a) Lignine, b) 3-aminophenyl boronic acid, c) N, B CQD, d) N, B CQD 20 μl HCl, e) N, B CQD 1000 μl HCl

**Figure S3**: 2D-HSQC NMR spectra of a) Kraft lignin, b) acidolysis components, and c) CQDs

**Table S1**: C1s bonds in kraft lignin

**Table S2**: C1s bonds at N, B, CQDs

**Table S3**: C1s bonds in N, B CQD 20 μl HCl

**Table S4**: C1s bonds in N, B CQD 1000 μl HCl

**Table S5**: The area of each sub-unit (S, G, H, and -OCH$_3$) and the linkages between (A, B, C) in the 2D HSQC NMR spectrum.

**Author information**

Corresponding authors:

**Jelena Papan Djaniš**, Faculty of Chemistry and Chemical Technology, University of Ljubljana, 1000 Ljubljana, Slovenia, Centre of Excellence for Photoconversion, Vinča Institute of Nuclear

Sciences, National Institute of the Republic of Serbia, University of Belgrade, 11351 Belgrade, Serbia (jelena.papandjanis@fkkt.uni-lj.si)

**Karolina Elzbieciak-Piecka**, Institute of Low Temperature and Structure Research Polish Academy of Sciences, Wroclaw(k.elzbieciak@intibs.pl)

Authors:

**Jelena Papan Djaniš**, Faculty of Chemistry and Chemical Technology, University of Ljubljana, 1000 Ljubljana, Slovenia, Centre of Excellence for Photoconversion, Vinča Institute of Nuclear Sciences, National Institute of the Republic of Serbia, University of Belgrade, 11351 Belgrade, Serbia (jelena.papandjanis@fkkt.uni-lj.si)

**Maja Szymczak**, Institute of Low Temperature and Structure Research Polish Academy of Sciences, Wroclaw (m.szymczak@intibs.pl)

**Jan Hočevar**, Faculty of Chemistry and Chemical Technology, University of Ljubljana, 1000 Ljubljana, Slovenia (jan.hocevar@fkkt.uni-lj.si)

**Jernej Iskra**, Faculty of Chemistry and Chemical Technology, University of Ljubljana, 1000 Ljubljana, Slovenia (jernej.iskra@fkkt.uni-lj.si)

**Boštjan Genorio**, Faculty of Chemistry and Chemical Technology, University of Ljubljana, 1000 Ljubljana, Slovenia (bostjan.genorio@fkkt.uni-lj.si)

**Darja Lisjak**, Department for the Synthesis of Materials, Jožef Stefan Institute, 1000 Ljubljana, Slovenia (darja.lisjak@ijs.si)

**Lukasz Marciniak,** Institute of Low Temperature and Structure Research Polish Academy of Sciences, Wroclaw (l.marciniak@intibs.pl)

**Karolina Elzbieciak-Piecka**, Institute of Low Temperature and Structure Research Polish Academy of Sciences, Wroclaw(k.elzbieciak@intibs.pl)

**Author Contributions**

The manuscript was written through contributions of all authors. All authors have given approval to the final version of the manuscript.

**Acknowledgment**:

The authors gratefully acknowledge the financial support from the Slovenian Research Agency (ARIS) – research core funding grants P1-0134, P2-0423 and P1-0418, research project J2-50061, and Young Researcher Grant to J.H.). K.E.P., M.Sz. and L.M. acknowledge the support from National Science Center Poland under UMO- 2023/05/Y/ST5/00013 project.

**References**:

[1]   D. D. S. Argyropoulos, C. Crestini, C. Dahlstrand, E. Furusjö, C. Gioia, K. Jedvert, G. Henriksson, C. Hulteberg, M. Lawoko, C. Pierrou, J. S. M. Samec, E. Subbotina, H. Wallmo, and M. Wimby., 'Kraft Lignin: A Valuable, Sustainable Resource, Opportunities and Challenges', *ChemSusChem*, vol. 17, no. 23, 2023, doi: 10.1002/cssc.202300492.

[2] V. K. Ponnusamy, D. D. Nguyen, J. Dharmaraja, S. Shobana, J. Rajesh Banu, J. G. Sartale, S. W. Chang, G. Kumar, 'A review on lignin structure, pretreatments, fermentation reactions and biorefinery potential', *Bioresour Technol*, vol. 271, 2019. doi: 10.1016/j.biortech.2018.09.070.

[3] S. S. Wong, R. Shu, J. Zhang, H. Liu, and N. Yan, 'Downstream processing of lignin derived feedstock into end products', *Chem Soc Rev*, vol. 49, no. 15, 2020. doi: 10.1039/d0cs00134a.

[4] H. Luo and M. M. Abu-Omar, 'Lignin extraction and catalytic upgrading from genetically modified poplar', *Green Chemistry*, vol. 20, no. 3, 2018, doi: 10.1039/c7gc03417b.

[5] Y. Xue, X. Qiu, Y. Wu, Y. Qian, M. Zhou, Y. Deng and Y. Li, 'Aggregation-induced emission: The origin of lignin fluorescence', *Polym Chem*, vol. 7, no. 21, 2016, doi: 10.1039/c6py00244g.

[6] W. Chen, C. Hu, Y. Yang, J. Cui, and Y. Liu, 'Rapid synthesis of carbon dots by hydrothermal treatment of lignin', *Materials*, vol. 9, no. 3, 2016, doi: 10.3390/ma9030184.

[7] L. Zhu, D. Shen, Q. Wang, and K. H. Luo, 'Green Synthesis of Tunable Fluorescent Carbon Quantum Dots from Lignin and Their Application in Anti-Counterfeit Printing', *ACS Appl Mater Interfaces*, vol. 13, no. 47, 2021, doi: 10.1021/acsami.1c16679.

[8] L. Zhu, D. Shen, Q. Liu, C. Wu, and S. Gu, 'Sustainable synthesis of bright green fluorescent carbon quantum dots from lignin for highly sensitive detection of Fe3+ ions', *Appl Surf Sci*, vol. 565, 2021, doi: 10.1016/j.apsusc.2021.150526.

[9] L. Zhu, H. Wu, S. Xie, H. Yang, and D. Shen, 'Multicolor lignin-derived carbon quantum dots: Controllable synthesis and photocatalytic applications', *Appl Surf Sci*, vol. 662, 2024, doi: 10.1016/j.apsusc.2024.160126.

[10] X. Yang, Y. Guo, S. Liang, S. Hou, T. Chu, J. Ma, X. Chen, J. Zhou and R. Sun, 'Preparation of sulfur-doped carbon quantum dots from lignin as a sensor to detect Sudan i in an acidic environment', *J Mater Chem B*, vol. 8, no. 47, 2020, doi: 10.1039/d0tb00125b.

[11] Y. Wang, Y. Liu, J. Zhou, J. Yue, M. Xu, B. An, C. Ma, W. Li and S. Liu, 'Hydrothermal synthesis of nitrogen-doped carbon quantum dots from lignin for formaldehyde determination', *RSC Adv*, vol. 11, no. 47, 2021, doi: 10.1039/d1ra05370a.

[12] L. Zhu, D. Shen, and K. Hong Luo, 'Triple-emission nitrogen and boron co-doped carbon quantum dots from lignin: Highly fluorescent sensing platform for detection of hexavalent chromium ions', *J Colloid Interface Sci*, vol. 617, 2022, doi: 10.1016/j.jcis.2022.03.039.

[13] L. Zhu, H. Wu, S. Xie, H. Yang, and D. Shen, 'Multicolor lignin-derived carbon quantum dots: Controllable synthesis and photocatalytic applications', *Appl Surf Sci*, vol. 662, 2024, doi: 10.1016/j.apsusc.2024.160126.

[14] S. Zhao, X. Chen, C. Zhang, P. Zhao, A. J. Ragauskas, and X. Song, 'Fluorescence Enhancement of Lignin-Based Carbon Quantum Dots by Concentration-Dependent and Electron-Donating Substituent Synergy and Their Cell Imaging Applications', *ACS Appl Mater Interfaces*, vol. 13, no. 51, 2021, doi: 10.1021/acsami.1c20648.

[15] L. Zhu, H. Wu, H. Yang, D. Shen, H. Hu, and M. Dou, 'Formation mechanism of lignin-derived carbon quantum dots: From chemical structures to fluorescent behaviors', *Bioresour Technol*, vol. 413, 2024, doi: 10.1016/j.biortech.2024.131490.

[16] H. Nawaz, X. Zhang, S. Chen, X. Li, X. Zhang, I. Shabbir, and F. Xu, 'Recent developments in lignin-based fluorescent materials', *Int J Biol Macromol*, vol. 258, 2024. doi: 10.1016/j.ijbiomac.2023.128737.

[17] T. Yang, Y. He, C. Wang, H. Bi, and G. Chen, 'Carbon quantum dots derived from lignin nanoparticles for dual UV-excited fluorescent anti-counterfeiting materials', *Int J Biol Macromol*, p. 140666, Feb. 2025, doi: 10.1016/j.ijbiomac.2025.140666.

[18] B. Zhang, Y. Liu, M. Ren, W. Li, X. Zhang, R. Vajtai, P. M. Ajayan, J. M. Tour, and L. Wang, 'Sustainable Synthesis of Bright Green Fluorescent Nitrogen-Doped Carbon Quantum Dots from Alkali Lignin', *ChemSusChem*, vol. 12, no. 18, 2019, doi: 10.1002/cssc.201901693.

[19] A. Haldar, P. Bhagwati, and J. Ekhe, 'Adsorptive Removal of Malachite Green using the Coke Obtained from Pyrolysis of Industrial Waste Lignin',*J Chem Biol Phys Sci*, vol. 6, no. 3 2016.

[20] R. K. Das and S. Mohapatra, 'Highly luminescent, heteroatom-doped carbon quantum dots for ultrasensitive sensing of glucosamine and targeted imaging of liver cancer cells', *J Mater Chem B*, vol. 5, no. 11, pp. 2190–2197, 2017, doi: 10.1039/c6tb03141b.

[21] Ł. Klapiszewski, M. Wysokowski, I. Majchrzak, T. Szatkowski, M. Nowacka, K. Siwińska-Stefańska, K. Szwarc-Rzepka, P. Bartczak, H. Ehrlich, and T. Jesionowski, 'Preparation and characterization of multifunctional chitin/lignin materials', *J Nanomater*, vol. 2013, 2013, doi: 10.1155/2013/425726.

[22] Y. Xu, T. Wang, Z. Chen, Y. Li, D. Huang, F. Guo, M. Wang, 'Hydrolysis of p-Phenylenediamine Antioxidants: The Reaction Mechanism, Prediction Model, and Potential

Impact on Aquatic Toxicity', *Environ Sci Technol*, vol 59, no 1, 2024, doi: 10.1021/acs.est.4c10227.

[23] Y. Li, M. Hu, K. Liu, S. Gao, H. Lian, and C. Xu, 'Lignin derived multicolor carbon dots for visual detection of formaldehyde', *Ind Crops Prod*, vol. 192, 2023, doi: 10.1016/j.indcrop.2022.116006.

[24] B. Wang, J. Yu, L. Sui, S. Zhu, Z. Tang, B. Yang, S. Lu, 'Rational Design of Multi-Color-Emissive Carbon Dots in a Single Reaction System by Hydrothermal', *Advanced Science*, vol. 8, no. 1, 2021, doi: 10.1002/advs.202001453.

[25] L. Ai, Z. Song, M. Nie, J. Yu, F. Liu, H. Song, B. Zhang, G. I. N. Waterhouse, S. Lu, 'Solid-state Fluorescence from Carbon Dots Widely Tunable from Blue to Deep Red through Surface Ligand Modulation', *Angew Chem Int Ed*, vol. 62, no. 12, 2023, doi: 10.1002/anie.202217822.

[26] M. Karlsson, J. Romson, T. Elder, Å. Emmer, and M. Lawoko, 'Lignin Structure and Reactivity in the Organosolv Process Studied by NMR Spectroscopy, Mass Spectrometry, and Density Functional Theory', *Biomacromolecules*, vol. 24, no. 5, 2023, doi: 10.1021/acs.biomac.3c00186.

[27] X. Liu, S. Zhao, X. Chen, X. Han, J. Zhang, M. Wu, X. Song, and Z. Zhang, 'The effect of lignin molecular weight on the formation and properties of carbon quantum dots', *Green Chemistry*, vol. 26, no. 6, 2024, doi: 10.1039/d3gc04694j.

[28] T. R. Kozmelj, E. Bartolomei, A. Dufour, S. Leclerc, P. Arnoux, B. Likozar, E. Jasiukaitytė-Grojzdek, M. Grilc, Y. Le Brech, 'Oligomeric fragments distribution, structure and

functionalities upon ruthenium-catalyzed technical lignin depolymerization', *Biomass Bioenergy*, vol. 181, 2024, doi: 10.1016/j.biombioe.2024.107056.

[29] S. Lin, C. Lai, Z. Huang, W. Liu, L. Xiong, Y. Wu, Y. Jin, 'Sustainable synthesis of lignin-derived carbon dots with visible pH response for $Fe^{3+}$ detection and bioimaging', *Spectrochim Acta A Mol Biomol Spectrosc*, vol. 302, 2023, doi: 10.1016/j.saa.2023.123111.

[30] M. L. Liu, B. Bin Chen, C. M. Li, and C. Z. Huang, 'Carbon dots: Synthesis, formation mechanism, fluorescence origin and sensing applications', *Green Chem*, vol. 21, 2019, doi: 10.1039/c8gc02736f.

[31] X. Li, S. Zhang, S. A. Kulinich, Y. Liu, and H. Zeng, 'Engineering surface states of carbon dots to achieve controllable luminescence for solid-luminescent composites and sensitive $Be^{2+}$ detection', *Sci Rep*, vol. 4, 2014, doi: 10.1038/srep04976.

[32] S. Zhu, Y. Song, X. Zhao, J. Shao, J. Zhang, and B. Yang, 'The photoluminescence mechanism in carbon dots (graphene quantum dots, carbon nanodots, and polymer dots): current state and future perspective', *Nano Res*, vol. 8, no. 2, pp. 355–381, 2015, doi: 10.1007/s12274-014-0644-3.

[33] M. Alafeef, I. Srivastava, T. Aditya, and D. Pan, 'Carbon Dots: From Synthesis to Unraveling the Fluorescence Mechanism', *Small*, vol. 20, no. 4, 2024, doi: 10.1002/smll.202303937.

[34] F. Yan, Z. Sun, H. Zhang, X. Sun, Y. Jiang, and Z. Bai, 'The fluorescence mechanism of carbon dots, and methods for tuning their emission color: a review', *Microchimica Acta*, vol. 186, no. 8, 2019, doi: 10.1007/s00604-019-3688-y.

[35] H. Ding, S. B. Yu, J. S. Wei, and H. M. Xiong, 'Full-color light-emitting carbon dots with a surface-state-controlled luminescence mechanism', *ACS Nano*, vol. 10, no. 1, pp. 484–491, 2016, doi: 10.1021/acsnano.5b05406.

[36] Z. L. Wu, Z. X. Liu, and Y. H. Yuan, 'Carbon dots: Materials, synthesis, properties and approaches to long-wavelength and multicolor emission', *J Mater Chem B*, vol. 5, no. 21, pp. 3794–3809, 2017, doi: 10.1039/c7tb00363c.

[37] J. Gan, L. Chen, Z. Chen, J. Zhang, W. Yu, C. Huang, Y. Wu, and K. Zhang, 'Lignocellulosic Biomass-Based Carbon Dots: Synthesis Processes, Properties, and Applications', *Small*, vol. 19, no. 48, 2023, doi: 10.1002/smll.202304066.

[38] J. Zhu, H. Shao, X. Bai, Y. Zhai, Y. Zhu, X. Chen, G. Pan, B. Dong, L. Xu, H. Zhang and H. Song, 'Modulation of the photoluminescence in carbon dots through surface modification: From mechanism to white light-emitting diodes', *Nanotechnology*, vol. 29, no. 24, 2018, doi: 10.1088/1361-6528/aab9d6.

[39] X. Zhao, S. Liang, Z. Li, X. Mao, M. Wang, X. Xie, and W. Gao, 'Lignin-Derived Carbon Dots with Triple Emission Peaks for Lighting Modules and Backlight Display', *ACS Appl Nano Mater*, vol. 6, no. 14, pp. 12893–12903, 2023, doi: 10.1021/acsanm.3c01633.

[40] G. E. LeCroy, F. Messina, A. Sciortino, C. E. Bunker, P. Wang, K. A. Shiral Fernando, and Y. P. Sun, 'Characteristic Excitation Wavelength Dependence of Fluorescence Emissions in Carbon "quantum" Dots', *Journal of Physical Chemistry C*, vol. 121, no. 50, pp. 28180–28186, 2017, doi: 10.1021/acs.jpcc.7b10129.

[41] Z. Gan, H. Xu, and Y. Hao, 'Mechanism for excitation-dependent photoluminescence from graphene quantum dots and other graphene oxide derivatives: Consensus, debates and challenges', *Nanoscale*, vol. 8, no. 15, pp. 7794–7807, 2016, doi: 10.1039/c6nr00605a.

**Graphical abstract:**

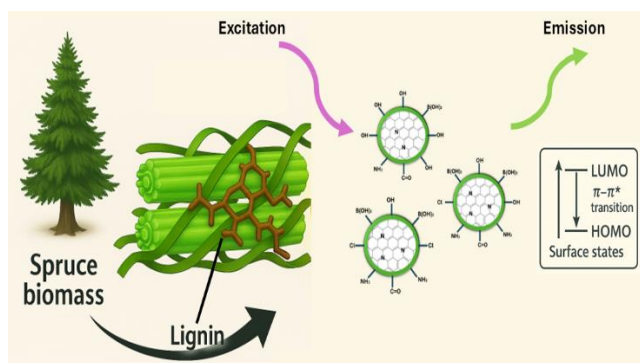

SYNOPSIS: Detailed analysis of LG CQD has shown that the acylolysis step is crucial for tuning the properties.